\documentclass[prl,twocolumn]{revtex4}
\usepackage{graphicx}

\begin{document}

\title{Statistical Signatures of Photon Localization}

\author{A.A. Chabanov} 

\author{M. Stoytchev}

\author{A.Z. Genack}

\affiliation{Physics Department, Queens College of CUNY, Flushing, NY 11367}

\date{April, 2000}

\maketitle

\textbf{The realization that electron localization in disordered systems (Anderson localization) \cite{Anderson} is ultimately a wave phenomenon \cite{Ioffe-Reagel,John84} has led to the suggestion that photons could be similarly localized by disorder \cite{John84}. This conjecture attracted wide interest because the differences between photons and electrons -- in their interactions, spin statistics, and methods of injection and detection -- may open a new realm of optical and microwave phenomena, and allow a detailed study of the Anderson localization transition undisturbed by the Coulomb interaction. To date, claims of three-dimensional photon localization have been based on observations of the exponential decay of the electromagnetic wave \cite{Genack1,Genack2,Wiersma1,Wiersma2,Vlasov} as it propagates through the disordered medium. But these reports have come under close scrutiny because of the possibility that the decay observed may be due to residual absorption \cite{Maret,Weaver,Yosefin}, and because absorption itself may suppress localization \cite{John84}. Here we show that the extent of photon localization can be determined by a different approach -- measurement of the relative size of fluctuations of certain transmission quantities. The variance of relative fluctuations accurately reflects the extent of localization, even in the presence of absorption. Using this approach, we demonstrate photon localization in both weakly and strongly scattering quasi-one-dimensional dielectric samples and in periodic metallic wire meshes containing metallic scatterers, while ruling it out in three-dimensional mixtures of aluminum spheres.}

In the absence of inelastic and phase-breaking processes, the ensemble average of the dimensionless conductance $\langle g\rangle\equiv\langle G\rangle/(e^{2}/h)$ is the universal scaling parameter \cite{Gang4} of the electron localization transition \cite{Anderson}. Here $\langle ...\rangle$ represents the average over an ensemble of random sample configurations, $G$ is the electronic conductance, $e$ is the electron charge, and $h$ is Planck's constant. The dimensionless conductance $g$ can be defined for classical waves as the transmittance, that is, the sum over transmission coefficients connecting all input modes $a$ and output modes $b$, $g\equiv\Sigma_{ab}T_{ab}$ (Ref.~13). In the absence of absorption, $\langle g\rangle$ not only determines the scaling of transmission quantities, such as $T_{ab}$ and $T_{a}=\Sigma_{b}T_{ab}$ that we will refer to as the intensity and total transmission, respectively, but it also determines their full distributions \cite{Rossum,Kogan,Beenakker}. In electronically conducting samples or in white paints, $\langle g\rangle\gg 1$ and Ohm's law holds, $\langle g\rangle=N\ell/L$, where $N$ is the number of transverse modes at a given frequency, $\ell$ is the transport mean free path, and $L$ is the sample length. But beyond the localization threshold, at $\langle g\rangle\approx 1$ (Refs.~12,17), the wavefunction or classical field is exponentially small at the boundary and $\langle g\rangle$ falls exponentially with $L$. Localization can be achieved in a strongly scattering three-dimensional sample with a sufficiently small value of $\ell$ (Ref.~2), or even in weakly scattering samples in a quasi-one-dimensional geometry of fixed $N$, once $L$ becomes greater than the localization length, $\xi = N\ell$ (Ref.~17). The latter corresponds to a wire in electronics or to a waveguide for microwave radiation.

In the presence of absorption, however, $\langle g\rangle$ cannot serve as a universal localization parameter because both the small value of $\langle g\rangle$ and its exponential scaling may simply reflect the effect of absorption. In this case, the decrease in $\langle g\rangle$ would represent a weakening rather than a strengthening of localization. Moreover, the distributions of the intensity and of the total transmission are affected by absorption and cannot be simply related to $\langle g\rangle$. Furthermore, it has been argued \cite{Shapiro} that $\langle g\rangle$ is not a natural scaling parameter because fluctuations in conductance are so large for localized waves that only the full conductance distribution or a parameter reflecting this distribution properly expresses the nature of transport. We find for microwave radiation that for $L\leq\xi$ the full distribution of the intensity and total transmission normalized to their ensemble averages, $s_{ab}=T_{ab}/\langle T_{ab}\rangle$ and $s_{a}=T_{a}/\langle T_{a}\rangle$, respectively, can be well expressed in terms of a single parameter, the variance of the normalized total transmission, $var(s_{a})$, even in strongly absorbing samples \cite{Marin1,Marin2}. This suggests that $var(s_{a})$ might serve as a localization parameter. Large values of $var(s_{a})$ would be expected for sharp spectra with widely spaced peaks that occur for localized waves. Further, because $var(s_{a})=2/3\langle g\rangle$ for $L\ll\xi, L_{a}$, where $L_{a}$ is the absorption length, and because the localization threshold occurs at $\langle g\rangle\approx 1$ in the absence of absorption, we make the conjecture that localization is achieved when $var(s_{a})\geq 2/3$. This localization condition may be expressed in a familiar form by defining a new localization parameter $g^{\prime}\equiv 2/3var(s_{a})$, which reduces to $\langle g\rangle$ in the absence of absorption in the limit $var(s_{a})\ll 1$. Localization is then achieved for $g^{\prime}\leq 1$ whether absorption is present or not.

In previous measurements of total transmission, $var(s_{a})$ was found to increase sublinearly with length for diffusive waves in strongly absorbing quasi-one-dimensional dielectric samples \cite{Marin1}. This raised the possibility that the value of $var(s_{a})$ might saturate with length, and that absorption might introduce a cut-off length for the renormalization of transport. Here we show that, though the presence of absorption leads to a decrease in $var(s_{a})$, this appropriately reflects a lessening of localization effects. The threshold for localization occurs at $g^{\prime}\approx 1$, and for smaller values, $g^{\prime}$ falls exponentially with length.

We now consider wave transport statistics in a quasi-one-dimensional geometry. Because $L$ is much greater than the transverse dimensions of a quasi-one-dimensional sample, energy injected at any point of the input is equally likely to emerge at any point of the output, and modes are completely mixed by the medium. As a result, statistical measurements at any point on the output surface of the sample or in any of its transmission modes yield identical results, and depend upon the sample geometry only through the ratio $\xi /L=\langle g\rangle$. For $N\gg 1$, scattering is locally three-dimensional and wave transport may be given a universal description.

\begin{figure}[!]
\includegraphics[width=\columnwidth]{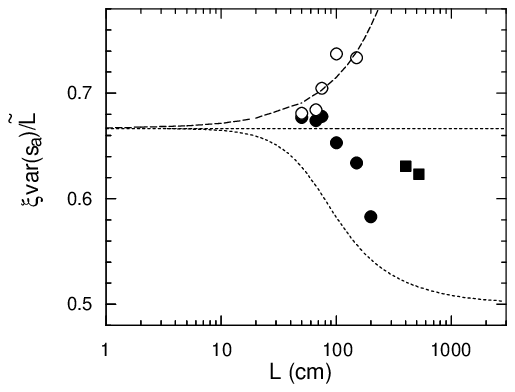}
\caption{Influence of absorption and localization, separately and together, on $var(s_{a})$ in random polystyrene samples. A semi-logarithmic plot of $\xi var(s_{a})/\tilde{L}$ is presented to illustrate the measured scaling of $var(s_{a})$ over a large range of $L$, as well as various theoretical predictions. The upper and lower short-dashed lines represent the two limits of diffusion theory: $L\ll\xi,L_{a}$ and $L_{a}\ll L\ll\xi$, respectively. The filled circles are obtained from measurements of total transmission, while the filled squares are obtained from measurements of intensity. The open circles are the results of an analysis that eliminates the effect of absorption, as explained in the text. The upper, long-dashed curve is a fit of these results to an expression incorporating the first-order localization correction to diffusion theory.}
\label{fig1}
\end{figure}

We will first consider the scaling of $var(s_{a})$ and its connection to localization in strongly absorbing weakly scattering dielectric samples contained in a copper tube. The role of absorption will be investigated by comparing these measurements to an analysis of the data that statistically eliminates the influence of absorption. Measurements are carried out in samples of loosely packed, 1.27-cm-diameter polystyrene spheres with a filling fraction of 0.52 within the frequency range 16.8-17.8 GHz. In these samples, $\ell\approx 5$ cm (Ref.~21), giving $\xi\approx 5$ m for a tube diameter of 5 cm. The exponential attenuation length due to absorption is $L_{a}=0.34\pm 0.02$ m (Ref.~19), and the diffusion extrapolation length, which gives an effective sample length for the statistics of transmission \cite{Ad}, $\tilde{L}=L+2z_{b}$, is $z_{b}\approx 6$ cm (Ref.~21). For $L\ll\xi, L_{a}$, diffusion theory gives $var(s_{a})=2\tilde{L}/3\xi$ (Refs.~14,15). This result is shown as the horizontal short-dashed line in Fig.~\ref{fig1}. Measurements of fluctuations in the spectrum of total transmission in ensembles of polystyrene samples give the results shown as the filled circles in Fig.~\ref{fig1}. These results indicate that $var(s_{a})$ increases sublinearly with length up to $L=2$ m, which was the largest length at which accurate measurements of the total transmission could be made.

Values of $var(s_{a})$ can also be obtained from measurement of the intensity $T_{ab}$. Our measurements of intensity and total transmission confirm the predicted relation between the moments of the normalized intensity and total transmission \cite{Kogan}, $\langle s_{ab}^{n}\rangle=n!\langle s_{a}^{n}\rangle$. This allows us to relate the variance of the normalized transmission to the variance of the normalized intensity, which is more readily measured in microwave experiments,

\begin{equation}
2var(s_{a})=var(s_{ab})-1.
\label{}
\end{equation}

\noindent Using this relation, we are able to extend our study of statistics in random waveguides to greater lengths. We measured transmitted field spectra in an ensemble of 2,000 polystyrene samples with the use of a Hewlett-Packard 8772C network analyzer. The calculated intensity spectra yield $var(s_{ab})$, which gives the corresponding values of $var(s_{a})$ using equation (1). Values of $var(s_{a})$ obtained in this way for $L\leq 2$ m agree to within 3\% with those shown as the filled circles in Fig.~\ref{fig1}. The results for $L>2$ m are shown in the figure as the filled squares. They indicate a more rapid, superlinear increase in $var(s_{a})$ relative to the data for $L\leq 2$ m.

In these measurements, the effect of developing localization and absorption are intertwined. In order to obtain the values of $var(s_{a})$ that would be measured in the absence of absorption, the field spectra are Fourier-transformed to give the response to a narrow gaussian pulse in the time domain. To compensate for losses due to absorption, the time-dependent field is multiplied by $\exp(t/2\tau_{a})$, where $t$ is the time delay from the incident pulse and $1/\tau_{a}$ is the absorption rate determined from measurements of the field correlation function with frequency shift \cite{Garcia1}. This new field is transformed back to the frequency domain. Intensity spectra and the distribution and variance of intensity are then computed. The intensity distributions are in excellent agreement with calculations for diffusive waves \cite{Rossum,Kogan}, which are described in terms of a single parameter $\langle g\rangle$. The values of $var(s_{a})$ found in this way are shown as the open circles in Fig.~\ref{fig1}. A fit of the leading order localization correction \cite{Rossum}, $var(s_{a})=2\tilde{L}/3\xi + 4\tilde{L}^{2}/15\xi^{2}$, to the data gives the upper long-dashed curve in Fig.~\ref{fig1} with $\xi=5.51\pm 0.18$ m and $z_{b}=5.25\pm 0.31$ cm. The results are consistent with independent determination of these parameters \cite{Lisyansky}. The difference between the open and filled circles represents the amount by which $var(s_{a})$ is reduced, and hence represents the extent to which localization is suppressed by absorption.

For diffusing waves, $var(s_{a})$ is predicted to fall from $2\tilde{L}/3\xi$ for $L\ll L_{a}$, to $\tilde{L}/2\xi$ for $L\gg L_{a}$ (Refs.~24,25), following the lower short-dashed curve in Fig.~\ref{fig1}. Notwithstanding the initial drop of $var(s_{a})$ from $2\tilde{L}/3\xi$, our measurements rise above this curve as a result of enhanced intensity correlation, as $L\to\xi$. At $L=5.2$ m, $var(s_{a})=0.6$, which is close to the critical value of 2/3.

\begin{figure}[!]
\includegraphics[width=\columnwidth]{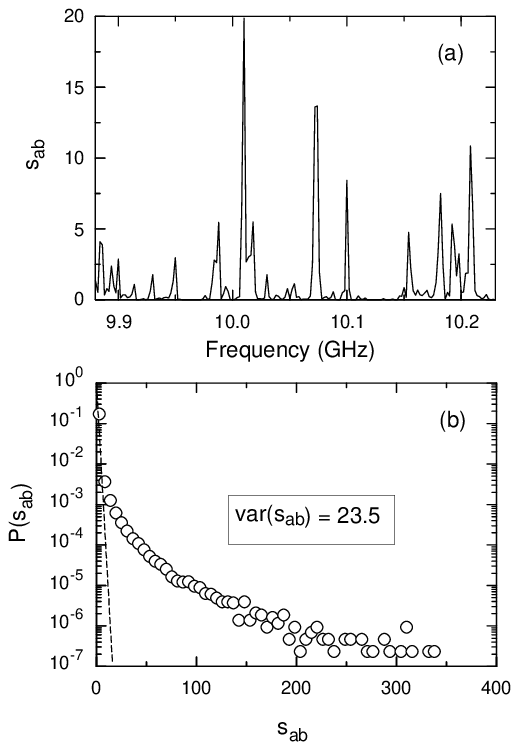}
\caption{Statistics of intensity in quasi-one-dimensional alumina samples. Alumina spheres (diameter $d_{a}=0.95$ cm, dielectric constant $\epsilon_{a}=9.8$) embedded in Styrofoam shells ($d_{S}=1.9$ cm, $\epsilon_{S}=1.05$) at an alumina volume fraction of 0.068 are contained in an 80-cm-long, 7.3-cm-diameter copper tube. A typical spectrum of the normalized intensity $s_{ab}$ (a) and the distribution $P(s_{ab})$ (b) near the first Mie resonance of the alumina spheres are presented. The sharp and narrow line spectra and giant fluctuations shown have been predicted for localized waves, and are unlike corresponding spectra and distributions in diffusive samples \cite{Garcia1}. The distribution $P(s_{ab})$ plotted on a semi-logarithmic scale is computed in an ensemble of 5,000 sample configurations within the frequency range 9.88-10.24 GHz, in which statistical parameters do not change substantially. The broken line shows the Rayleigh distribution.}
\label{fig2}
\end{figure}

\begin{figure}[!]
\includegraphics[width=\columnwidth]{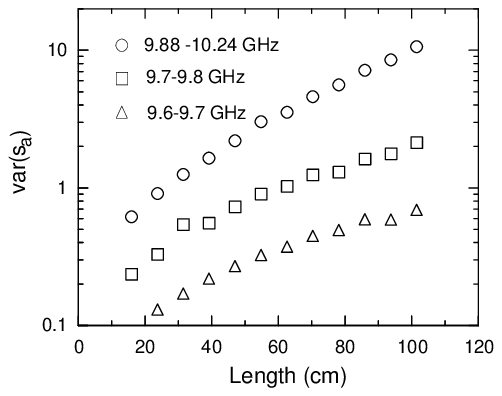}
\caption{Scaling of $var(s_{a})$ in alumina samples. The values of $var(s_{a})$ averaged over the indicated frequency intervals are obtained using equation (1). Above a value of the order of unity, $var(s_{a})$ increases exponentially. In the interval 9.88-10.24 GHz, $var(s_{a})\approx\exp(L/L_{exp})$, with $L_{exp}\approx 42$ cm.}
\label{fig3}
\end{figure}

To study the statistics of $s_{a}$ for localized waves, we examine fluctuations of intensity in strongly scattering quasi-one-dimensional samples of alumina (Al$_{2}$O$_{3}$) spheres. A typical spectrum of $s_{ab}$ obtained near the first Mie resonance of the spheres is shown in Fig.~\ref{fig2}a. The sharp peaks in $s_{ab}$ result from resonant transmission through localized photonic states in the medium. The distribution function $P(s_{ab})$ calculated for an ensemble of 5,000 samples is shown in Fig.~\ref{fig2}b and compared to the Rayleigh distribution. The measured distribution is remarkably broad, with $var(s_{ab})=23.5$ and fluctuations greater than 300 times the average value. The scaling of $var(s_{a})$ determined using equation (1) at a number of frequencies is shown in Fig.~\ref{fig3}. We find $var(s_{a})$ increases exponentially once it becomes of order of unity, as expected for a localization parameter.

\begin{figure}[!]
\includegraphics[width=\columnwidth]{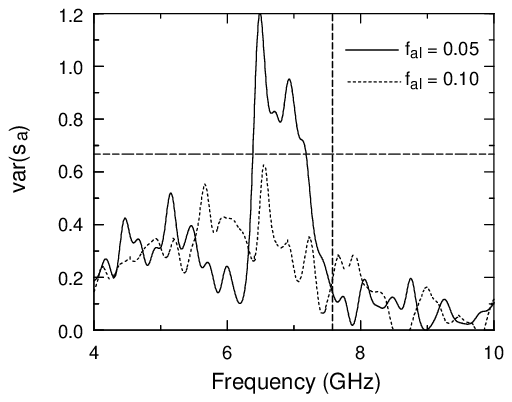}
\caption{$Var(s_{a})$ versus frequency in a wire-mesh photonic crystal containing metal scatterers. The photonic crystal is a simple cubic lattice made up of copper wires, with a lattice constant of 1 cm. The lattice has 8 unit cells along each side. It is enclosed in a section of a square waveguide and filled with mixtures of 0.47-cm-diameter aluminum and Teflon spheres, the latter used to dilute the aluminum scatterers. Measurements of intensity are carried out in 200 sample configurations, and $var(s_{a})$ is obtained using equation (1). The broken vertical line indicates the position of the band edge in a periodic structure filled only with Teflon spheres. At an aluminum sphere volume fraction ($f_{\rm al}$) of 0.05, $var(s_{a})$ is markedly higher near the edge, rising above the localization threshold of 2/3 shown as the broken horizontal line.}
\label{fig4}
\end{figure}

The availability of a measurable localization parameter makes it possible to determine the existence and the extent of localization in a variety of samples. This is illustrated in measurements of localization in periodic metallic wire meshes containing metallic scatterers. John has proposed that photon localization could be achieved by introducing disorder in a periodic structure possessing a photonic bandgap \cite{John87}. Measurements of transmission in the wire-mesh photonic crystal show a low-frequency gap \cite{Soukoulis}, which fills in as the scatterer density is increased \cite{Marin3}. But such measurements leave open the question of whether the radiation is localized. To answer this question, we obtain $var(s_{a})$ for two concentrations of aluminum spheres shown in Fig.~\ref{fig4}. At a volume fraction of aluminum spheres of 0.05, a window of localization is found, in which $var(s_{a})\geq 2/3$. At twice this aluminum fraction, the reduced values of $var(s_{a})$ indicate that wave propagation is diffusive.

We have also used measurements of $var(s_{a})$ to examine the claim that localization can be achieved in three-dimensional samples of metal spheres at various concentrations \cite{Genack1,Genack2,Birman,Kirkpatrik}. We find that in samples of 0.47-cm-diameter aluminum spheres of length $L=8.2$ cm and diameter $d=7.5$ cm, with various volume fraction from 0.1 to 0.475, $var(s_{a})$ never rises above the localization threshold of 2/3. A maximum value of 0.29 is reached at a concentration of 0.45. Thus three-dimensional localization is not achieved in these aluminum samples. These results demonstrate that $var(s_{a})$ is a powerful guide in the search for and characterization of photon localization.

\begin{acknowledgments}
We thank P.W.~Brouwer and E.~Kogan for stimulating discussions. This work was supported by NSF.
\end{acknowledgments}

\line(1,0){228}

\end{document}